# 3UCubed: The IMAP Student Collaboration CubeSat Project

**Sanjeev Mehta, Noé Lugaz, Lindsay Bartolone, Marc Lessard**

Jenna Burgett, Kelly Bisson, Luke Bradley, Jeffrey Campbell, Mathieu Champagne, Alex Chesley, Jeff Grant, Hanyu Jiang, Jared King, Emily McLain, Matthew Rollend, Shane Woods

University of New Hampshire

Space Science Centre, 8 College Rd, Durham, NH 03824; (603) 862-2732

sanjeev.mehta@unh.edu

**Laura Peticolas, Lynn Cominsky, Garrett Jernigan, Jeffrey Reedy, Doug Clarke**

Sabrina Blais, Erik Castellanos-Vasquez, Jack Dawson, Erika Diaz Ramirez, Walter Foster, Cristopher Gopar Carreno, Haley Joerger, Onasis Mora, Alex Vasquez

Sonoma State University

1801 East Cotati Ave, Rohnert Park, CA 94928

**Marcus Alfred, Sonya Smith, Charles Kim**

Carissma McGee, Ruth Davis, Myles Pope, Taran Richardson, Trinity Sager, Avery Williams, Matthew Gales, Wilson Jean Baptiste, Tyrese Kierstdet, Oluwatamilore Ogunbanjo

Howard University

2400 6th St NW, Washington, DC 20059

## ABSTRACT

The 3UCubed project is a 3U CubeSat being jointly developed by the University of New Hampshire, Sonoma State University, and Howard University as a part of the NASA Interstellar Mapping and Acceleration Probe (IMAP)[1] student collaboration. This project consists of a multidisciplinary team of undergraduate students from all three universities. The mission goal of the 3UCubed is to understand how Earth's polar upper atmosphere ('the thermosphere' in Earth's auroral regions) responds to particle precipitation and solar wind forcing and internal magnetospheric processes.

3UCubed includes two instruments with rocket heritage to achieve the science mission: an ultraviolet photomultiplier tube (UV-PMT) and electron retarding potential analyzer (ERPA). The spacecraft bus consists of the following subsystems–Attitude Determination and Control, Command and Data Handling, Power, Communication, Structural, and Thermal.

Currently, the project is in the post-PDR stage, starting to build and test engineering models to develop a FlatSat prior to critical design review in 2023. The goal is to launch at least one 3U CubeSat a to collect science data close to the anticipated peak of Solar Cycle 25 around July 2025.[2] Our mother mission– IMAP is also projected to launch in 2025, which will let us jointly analyze the science data of the main mission, providing the solar wind measurements and inputs to the magnetosphere with that of 3UCubed, providing the response of Earth's cusp to these inputs.

## INTRODUCTION

The 3UCubed project is a student-focused CubeSat collaboration effort between the University of New Hampshire, Sonoma State University, and Howard University, associated with NASA's Interstellar Mapping and Acceleration Probe (IMAP) mission. 3UCubed stands for 3 Universities; 3U CubeSats; Upwelling, Uplifting Undergraduate. 3UCubed has several main objectives: education, diversity, and science. The educational objective is to provide undergraduate students experience in space physics and engineering by providing opportunities in all stages of a CubeSat mission life cycle, from preliminary design to data analysis. This is supported by a strong recruitment and retention program to help diversify space engineering and space science and a dedicated program evaluation component. The science goal is to advance our understanding of Earth's upper atmosphere in the polar low Earth orbit (LEO), a region with numerous satellites and where the atmospheric density strongly affects satellite orbits and durations.

## MISSION OBJECTIVE

The education objective involves having undergraduate students actively participate in all phases of mission development, including its design. In addition, the goal is to improve diversity in STEM by having at least 50% of the participants being members of underrepresented groups in STEM, including women, Hispanic and African American students.

The science objective aims to specifically address how Earth's polar upper atmosphere (the thermosphere) in the auroral cusp regions responds to particle precipitation and varying conditions associated with solar wind forcing magnetospheric processes. To accomplish this science objective 3UCubed includes two instruments with rocket heritage. One instrument is the UV-PMT, which measures neutral atomic oxygen emissions to characterize (via remote-sensing) the most common neutral constituent in Earth's upper atmosphere. The second instrument is ERPA, which measures energetic precipitating electrons with energies between >10 and 150 electron-volts (eV). While the emphasis of the mission is on the cusp region, the instruments will measure UV emissions and precipitating electrons in both Earth's auroral and cusp regions. These 3UCubed measurements are combined with measurements from IMAP of the interplanetary plasma and magnetic field conditions to answer two science questions (SQ): "SQ1 How is the upwelling of the polar thermosphere affected by particle precipitation" and "SQ2 How does spatio-temporal variations in the polar thermosphere reflect variations in the solar wind forcing and magnetospheric conditions?"

## INSTRUMENTS

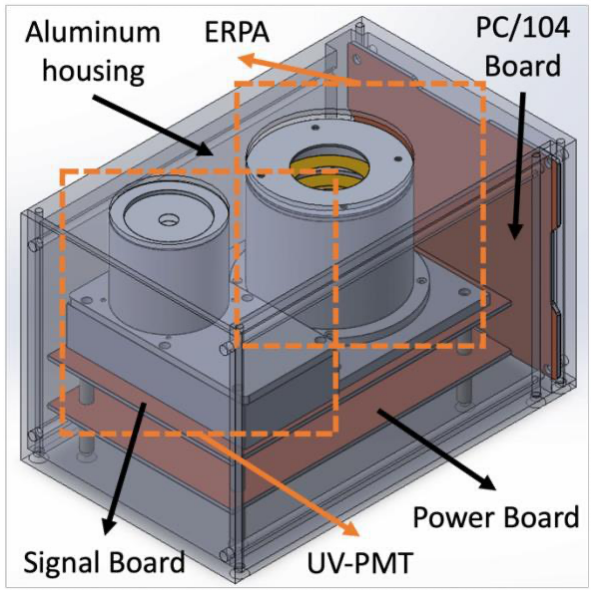

**Figure 1: Updated combined instrument design for 3UCubed**

### Electron Retarding Potential Analyzer–ERPA

The ERPA has been flown on multiple suborbital rocket missions: SERSIO (2004), ROPA (2007), SCIFER2 (2008), ACES (2009), MICA (2012), RENU2 (2015), KiNET-X (2021), and CREX-2 (2021). The ERPA as flown on rockets measured the distribution of electron energy ~0 to 3 eV, with a resolution of 0.06 eV at 2 ms cadence.[3]

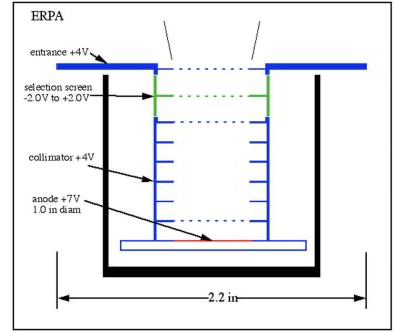

**Figure 2: Diagram showing internal structure of ERPA as flown on suborbital rockets**

The ERPA on 3UCubed will measure the flux of suprathermal precipitating electrons of energies from >10 to 150 eV in discrete steps. It measures the distribution of electron energy by sweeping or stepping a retarding potential across a selection screen of very fine conductive mesh. As the voltage steps across the selection screen, electrons with energies less than the retarding potential are rejected, while those with higher energies enter the instrument's optics and are recorded by the detector anode. The current collected by the anode is measured by a low noise electrometer circuit. The entrance face of the analyzer is held at a fixed +15V relative to payload ground to accelerate electrons into the entrance in the presence of a negative payload potential. Issues arising due to gyroradius effects are mitigated by the small dimensions of the instrument (roughly 5 cm in diameter) and ensuring that the instrument is oriented within ±15° to the magnetic field (which drives the ADCS requirements). The electrons from the external plasma are accelerated into the ERPA by the +15V bias, but are then decelerated after crossing the entrance screen as they approach the selection screen. The selection screen steps from -10 to -150 V in five steps of 100 ms duration. Once an electron with sufficient energy passes through the selection screen, it is accelerated toward the anode.[4] The ERPA as flown on past missions to measure ambient thermal ionospheric electrons in the energy range from a small fraction of an eV to 3 eV with a resolution of 0.06 eV. Since we expect the CubeSat to have a spacecraft potential (due to spacecraft charging) of up to 10 eV, the ERPA on 3UCubed has been modified to measure suprathermal electrons in the 10 to 150 eV range. A small modification on the selection screen voltage has been designed to step from -10 to -150 V instead on the existing +2 to -2 V to accommodate the same electronics with PC/104 standard for CubeSats. The accumulation time has been increased from 2 msec to 100 msec to account for the lower fluxes at suprathermal energies as compared to thermal.

### *Ultraviolet Photomultiplier Tube–UV-PMT*

The UV-PMT has also been flown on multiple rocket missions. It was first flown on SCIFER2 (2008) and has since been flown on RENU (2012) and RENU2 (2015) with an updated design and has successfully measured UV emissions of Oxygen at 130.4 nm and 135.6 nm.

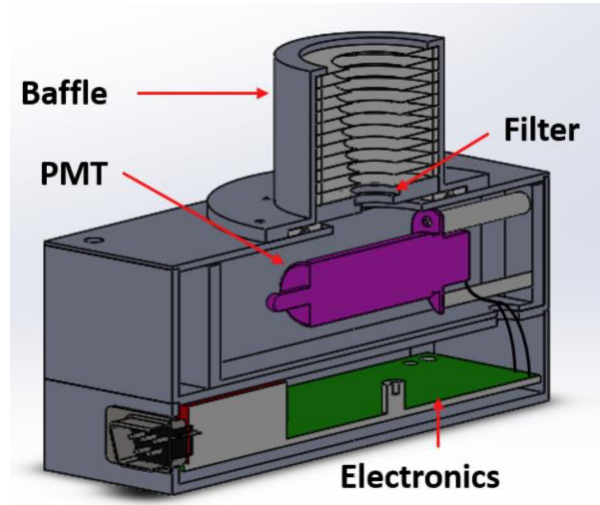

**Figure 3: CAD model of UV-PMT as flown on suborbital rockets**

The UV-PMT will detect emissions of neutral atomic oxygen from two common oxygen spectral lines in the auroral spectrum, OI 130.4 nm and OI 135.6 nm. The UV-PMT is a simple instrument built around Hamamatsu R10825 PMT. Photons enter the PMT through the baffled field of view (FoV) and filter, where they strike the surface of the photocathode and eject a photoelectron, which then cascades through a series of electrodes (dynodes) before reaching the anode. The current out of the anode is used to determine photon flux. The CsI photocathode and $MgF_2$ detector window allows an effective response from 115 to 195 nm, with a peak response at 130 nm. A $CaF_2$ filter from Pelham Research Optical, LLC is placed in front of the detector to exclude contamination from hydrogen Lyman-α emissions. A black-anodized, knife-edge baffle restricted the FoV to a 12.5° full angle cone and reduced glare from unwanted sources.[5] The only modifications as compared to the versions successfully flown on rocket is a reorganization of the electronics board to fit the CubeSat form factor. Undergraduate students have been working on the new board layout and schematics and the modified CAD for the two instruments.

### ORBIT

To achieve our science goal of characterizing the cusp an ideal orbit would be a high latitude orbit frequently passing the cusp, a polar orbit between 325 and 650 km with an inclination of 94°, RAAN of 50° and precession rate of 0.5°/day to ensure entire cusp is covered over the lifetime of spacecraft. LEO sun-synchronous orbit close to altitude of 500 km and LTAN close to noon-midnight is one of the acceptable orbits.

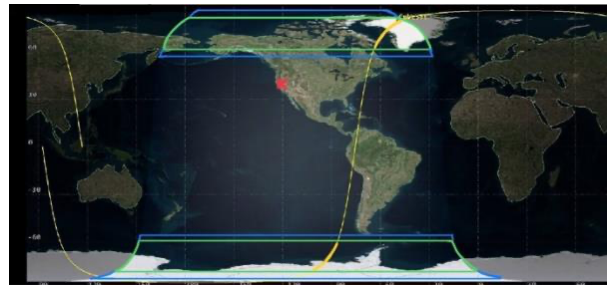

**Figure 4: Orbital simulation from STK showing yellow line as orbit of spacecraft, red 'x' as ground station at SSU, and green enclosed region as region of interest where cusp usually moves**

### SPACECRAFT BUS

The spacecraft bus consists of the following subsystems: Attitude Determination and Control, Command and Data Handling, Power, Communication, Structural, and Thermal.

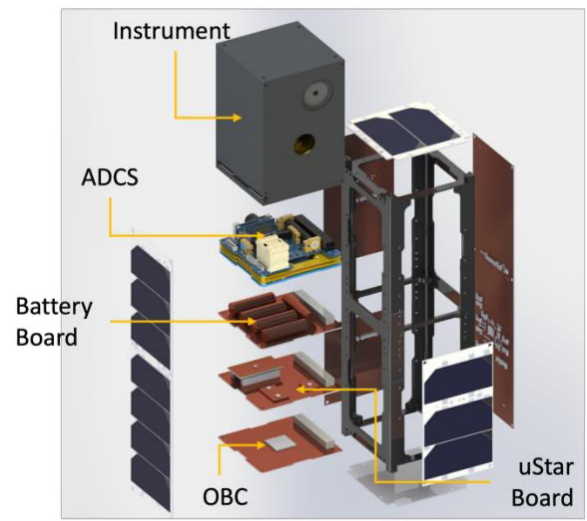

**Figure 5: Exploded view of 3UCubed CAD model**

### *Attitude Determination and Control System–ADCS*

To successfully measure the precipitating electrons the instrument's FoV should be aligned within ±15° with respect to field lines towards the zenith direction in the cusp region (60-85° GLAT). The instrument orientation requirement sets a unique ADCS requirement for our mission of acquiring attitude knowledge within ±5° accuracy and controlling the attitude within ±7.5° accuracy with respect to the field lines. We have considered passive magnet attitude control system to align the instruments FoV with magnetic field lines, but simulations show the CubeSat 'flipping' and 'rocking' at and around the poles makes it difficult to meet the ADCS requirements in the cusp region. A COTS integrated ADCS– CubeADCS Y-momentum from CubeSpace is chosen to meet these requirements. CubeADCS Y-momentum consists of CubeComputer– microcontroller with integrated magnetometer, 10 coarse sun sensors, and nadir sensor for attitude determination; 3-axes magnetorquers, and Y-axis momentum wheel controlling pitch for attitude control. The system also includes ADCS software which can be modified as per needs.

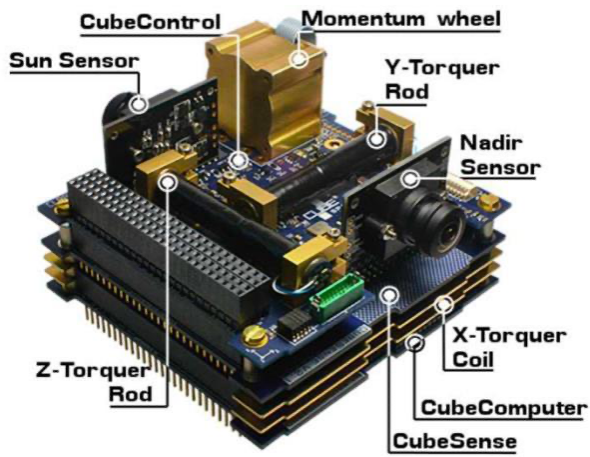

**Figure 6: CubeADCS Y-momentum COTS system from CubeSpace, from datasheet**

Upon deployment, CubeADCS provides functionality to detumble the spacecraft into a stable spin and then to absorb the rotation of the spacecraft into a momentum wheel. This momentum wheel provides good disturbance rejection which makes it possible to further keep the satellite 3-axis stable using magnetic control. The spacecraft can do pitch maneuvers in the orbital plane by controlling the momentum wheel's speed, this pitch control helps align the instrument with the magnetic field lines in the cusp region.

### Command and Data Handling–CD&H

The CD&H system consists of an on-board computer (OBC) that is being developed using an ARM microcontroller. An ARM microcontroller has been developed for the Sun Slicer instrument (selected for "Honey, I Shrunk the NASA Payload" mission for

Lunar payload)[6] by mentors for 3UCubed, a similar low-power ARM microcontroller will be used as the OBC for 3UCubed. The OBC is responsible for controlling all subsystems–scheduling and prioritizing tasks for subsystem level microcontrollers. The OBC for 3Ucubed is under development; the OBC will be fitted on a single PC/104 PCB, fitting in ~0.2U. The OBC will be equipped with flash memory, random access memory, redundant micro-SD cards for data storage, watch-dog timer, real-time clock, and fault protection. Various interfaces for instruments and subsystems like I2C, SPI, UART, USB are available. Flight software for OBC is being developed building upon SSU's past CubeSat and PocketQube– EdgeCube[7] and T-LogoQube[8] mission; the OBC's firmware will be programmed in both Logo and C/C++ languages.

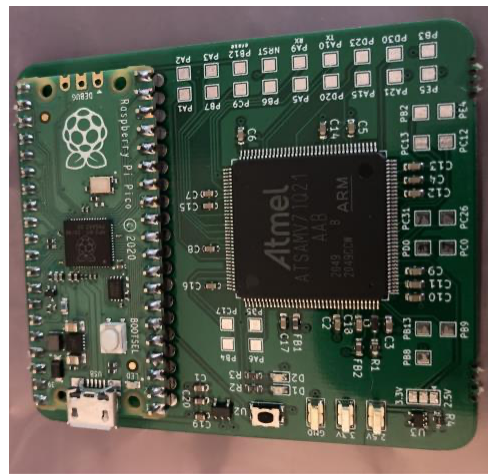

**Figure 7: Picture of ARM microcontroller designed for Sun Slicer instrument**

### Power

Power system consists of solar panels, electrical power system (EPS), and battery pack. COTS solar panels from EnduroSat comprising highly efficient Azur Space 3G30A triple junction solar cells with up to 30% efficiency, two of 3U, two of 1.5U, and three of 1U will be body mounted and will cover the entire body of the spacecraft except the instruments (1.5 U) and nadir sensor (0.5 U) FoV. The EPS consists of a low power microcontroller–uStar (previously used on SSU's EdgeCube) controlling power level operational modes, maximum power point tracking (MPPT) over voltage and current protection (OVP and OCP), power conditioning regulators, temperature monitors, and kill switches. The EPS operates in 3 modes– Full, Normal and Safe, depending on the battery charge state. Duty cycle of subsystems is defined to ensure that battery's state of charge is never completely depleted and essential subsystems like CD&H can always remain on.

The battery pack is built upon past design used for SSU's EdgeCube of 3-cell configuration; the battery pack for 3UCubed is being modified to accommodate 5 nickel metal hydride (Ni-MH) cells with an energy capacity of 16.5 W-hr, meeting our design requirements. The battery pack has integrated temperature monitors and heaters. A power budget for all three operational modes has been analyzed, resulting in positive power margin for all operational modes.

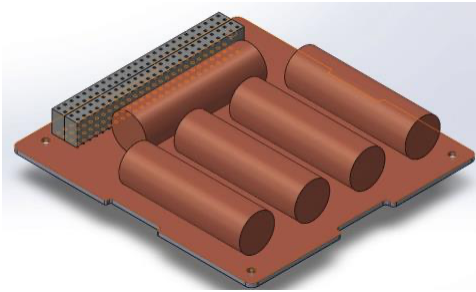

**Figure 8: CAD model of 5-cell battery pack based on SSU 3-cell design**

*Communication*

The communication system consists of 2 radios– primary and backup (beacon) radio, shared (with the power system) microcontroller–uStar, and a planar antenna. The communication system operates in amateur UHF band–435-438 MHz. The primary radio transmits all data collected on the spacecraft from the instruments and housekeeping, with transmitter power of 1 W and data rate; the backup radio is mainly used for beacon, and can downlink housekeeping data at low data rate (in case of fault with the primary radio), with transmitter power of 0.5 W. The uStar, which is the shared microcontroller with power system, controls both radios, collects data from other subsystems, and generates packets.

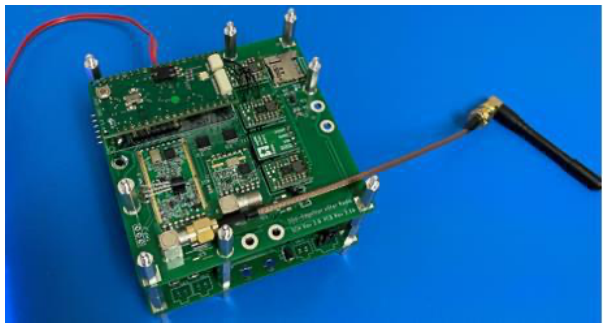

**Figure 9: SSU's uStar board consisting of 2 radios, shared microcontroller, and voltage regulators**

A planar antenna to operate in 435-438 MHz frequency range has been developed by Dr. John Doty of Noqsi Aerospace building upon directional discontinuity ring radiator (DDRR) antenna technology. The planar antenna has a maximum gain of -0.5 dBi, and the antenna can be easily tuned by changing the capacitance of a variable capacitor to tune the impedance to 50Ω at required frequency in 430-440 MHz range.

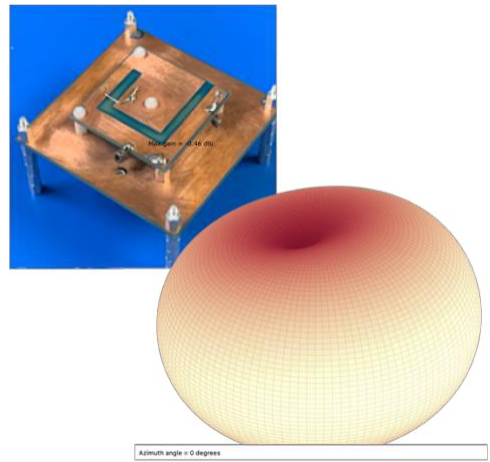

**Figure 10: (Left) Planar UHF antenna, (Right) radiation pattern of antenna from simulations**

A ground station used for past missions exists as SSU. The ground has 2 yagi antennas– a narrow beam, high power antenna, and a wide beam low gain antenna; both antenna are used with a common radio and front-end circuit with transmitter power of 10W. A similar ground station setup is planned to be built at UNH and HU following the SSU designs.

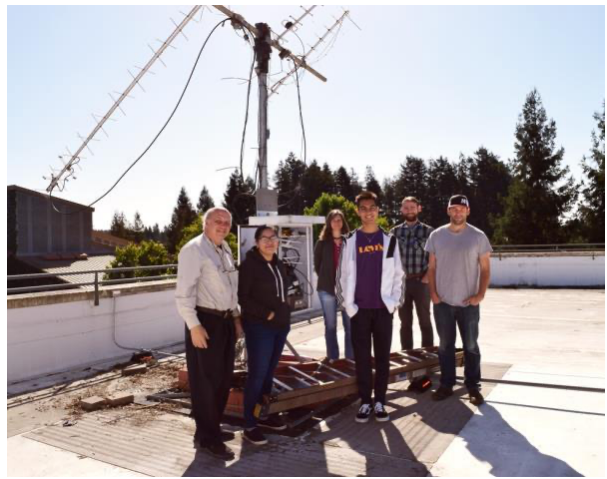

**Figure 11: Ground station setup at SSU with students and mentors**

A link margin has been analyzed, with sufficient link margin available for both uplink and downlink to complete link at 15° elevation for 500 km Sun-synchronous orbit.

### Structure

The COTS 3U structure from EnduroSat is chosen due its lightweight modular design, ease of assembly and the availability of wide openings that accommodate the 45° of unobstructed FoV required by the instrument. Additionally, the solar panels can be easily installed. The minimalistic and modular design of the structure comes as PC/104 base with rods, frame, mid supports, top plate, and securing hardware. The structure is fully compliant with CubeSat standards, comes with separation switches and optional roller switches on the rails. The structure is made of hard anodized space-grade 6061 aluminum, brass mounting rods for efficient heat conduction, and 304 stainless-steel securing hardware with a total mass of 340 g. The structure has passed vibration, shock, thermal-vacuum qualification test per NASA GEVS GSFC-STD-7000 standard.

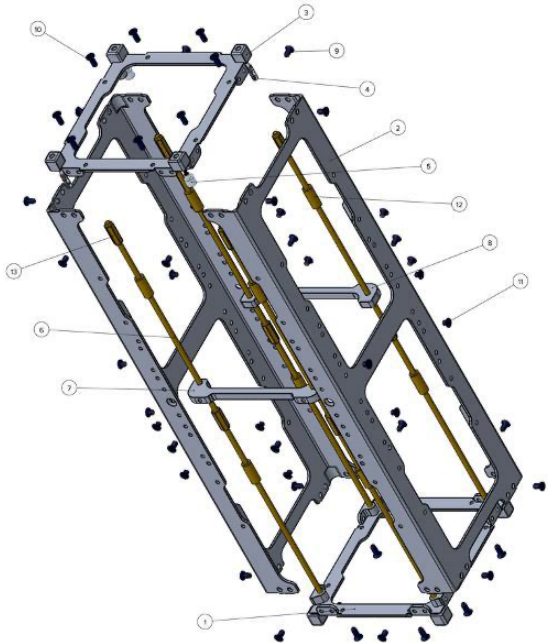

**Figure 12: Exploded view EnduroSat 3U Structure, from datasheet**

### Thermal

Preliminary thermal simulations have been performed in ANSYS by a team of students from Howard University and the University of New Hampshire to evaluate steady-state thermal performance in sunlit side (hot case) and transient eclipse (cold case). Preliminary results show that the temperature range of the spacecraft is within the operational temperature range of components in the spacecraft. A detailed transient thermal simulation in Thermal Desktop is being developed to evaluate thermal performance over the orbit.

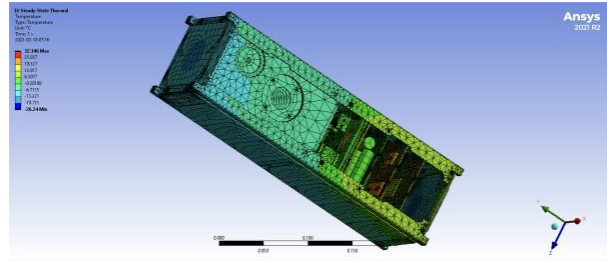

**Figure 13: ANSYS steady-state thermal simulation results for sunlit case**

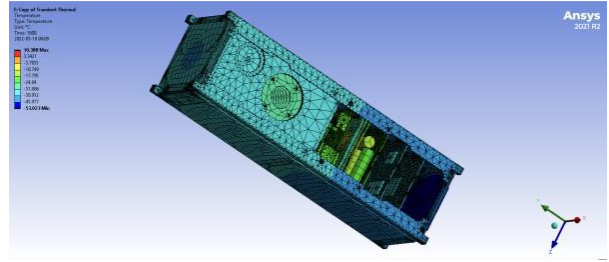

**Figure 14: ANSYS transient thermal simulation results for eclipse case**

### SUMMARY AND STATUS

In March 2022, we completed our preliminary design review, in addition, 3UCubed has been selected for launch on the 13th round of CubeSat Launch Initiative (CSLI). As a result, we are currently in the process of building prototypes and engineering models, to prepare a FlatSat. Some parts are being updated to comply with CubeSat PC/104 form factor.

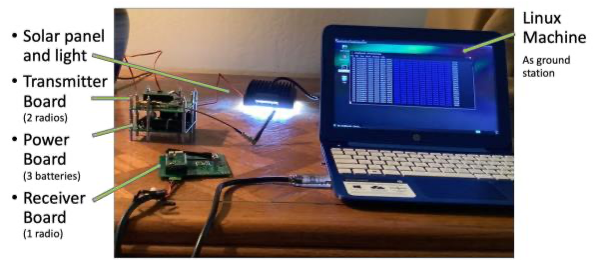

**Figure 15: SSU's EdgeCube spacecraft bus, currently being tested as FlatSat by the 3UCubed students**

### Acknowledgments

This work is supported by a grant from NASA as part of IMAP Student Collaboration, grant number 80NSSC20K1110.

The authors of this paper would like to thank all who contributed to the 3UCubed mission, all the students current and graduated: ***Howard University***: Oshione Adams - Electrical Engineering, Kelsy Coston -

Mechanical Engineering, Ruth Davis - Mechanical Engineering, Matthew Gales - Mechanical Engineering, Wilson Jean Baptiste - Physics and Astronomy – Graduate Student, Tyrese Kierstdet - Electrical Engineering, Oluwatamilore Ogunbanjo - Computer Engineer, Carissma McGee – Physics, Myles Pope - Physics, Taran Richardson - Physics, Trinity Sager - Physics, Avery Williams - Electrical Engineering; *Sonoma State University*: Sabrina Blais - Biochemistry and Theatre Arts, Erik Castellanos-Vasquez - UCLA, Aerospace Engineering, Graduated SSU, Jack Dawson - Physics, Erika Diaz Ramirez - Computer Science, Walter Foster - Electrical Engineering, Cristopher Gopar Carreno - Electrical Engineering, Haley Joerger - Computer Science, Onasis Mora - Physics, Alex Vasquez - Physics and Astronomy, Graduated SSU; *University of New Hampshire*: Jenna Burgett - Physics Graduate Student, Kelly Bisson - Engineering Physics, Luke Bradley - Electrical Engineering, Jeffrey Campbell – Graduated Engineering Physics, Mathieu Champagne - Mechanical Engineering, Alex Chesley – Graduated Mechanical Engineer, Jeff Grant - Mechanical Engineering, Hanyu Jiang – Graduated Engineering Physics, Jared King - Computer Science, Emily McLain - Engineering Physics, Matthew Rollend - Electrical Engineer, Shane Woods - Computer Science, and mentors: Dr. Doug Clarke, Dr. John Doty, Brian Silverman, and Gary Stofer. The team would also like to specially thank Dr. Garrett Jernigan for all his contributions to EdgeCube, T-LogoQube, and further bringing his experience and expertise to 3UCubed.

Further information about the project can be found at: https://imap.princeton.edu/engagement/student-collaboration